\documentclass[a4paper,11pt]{article}
\usepackage{jinstpub} 
\usepackage{placeins}

\title{\boldmath GEANT4-Based Comparative Study of Detector Response in Reduced and Fully Instrumented Silicon-Tungsten Sampling Calorimeter}

\author[a,1]{A. Laha,\note{Corresponding author.}}
\author[a]{S. Choudhury,}
\author[b,c]{S. Muhuri,}
\author[a,d]{M. Mondal}
\affiliation[a]{Nuclear and Particle Physics Research Centre, Department of Physics, Jadavpur University,\\Kolkata - 700032, India}
\affiliation[b]{Variable Energy Cyclotron Centre, 1/AF Bidhan Nagar, Kolkata, 700064, India}
\affiliation[c]{Homi Bhabha National Institute, Anushakti Nagar, Mumbai, 400094, India}
\affiliation[d]{School of Studies in Environmental Radiation and Archaeological Sciences, Jadavpur University,\\Kolkata 700032, India}

\emailAdd{abhijeetlaha41@gmail.com}
\emailAdd{subikash.choudhury@gmail.com}

\abstract{A GEANT4-based simulation study of a silicon--tungsten (Si--W) sampling calorimeter is presented to investigate electromagnetic shower development and detector response in the energy range of 1--120 GeV. Two detector configurations are considered: a reduced sequential-sampling prototype (Configuration-I), where a single silicon detector layer is placed downstream of progressively increasing tungsten absorber depths to sample the longitudinal shower evolution, and a fully instrumented sampling calorimeter (Configuration-II) consisting of twenty silicon layers interleaved with tungsten absorbers.

The detector response is studied in terms of longitudinal shower development, shower maximum position, energy deposition, detector linearity, and calorimetric performance. Configuration-II provides simultaneous longitudinal sampling and allows event-by-event reconstruction of the deposited energy, yielding the expected linear response and energy resolution. Configuration-I reproduces the average longitudinal shower profiles and exhibits a strong linear correlation with the integrated shower response of Configuration-II. Although the absence of event-by-event multi-layer sampling prevents a direct determination of calorimetric energy resolution in Configuration-I, the observed correlation between two configurations is used to relate the response of the reduced prototype to that of the fully instrumented calorimeter and to estimate the corresponding calorimetric performance.

The results demonstrate that the reduced prototype successfully reproduces the main characteristics of electromagnetic shower development observed in the fully instrumented detector, establishing its usefulness for detector characterization and test-beam studies when the construction of a complete prototype is limited by available resources.
}


\begin{document}
\maketitle
\flushbottom

\section{Introduction}
Electromagnetic calorimeters based on silicon--tungsten (Si--W) sampling technology are widely used in high-energy physics for the precise measurement of electromagnetic (EM) showers produced by charged particles and photons \cite{CALICE_2008, CALICE_2009, Awes_2021}. Their performance relies critically on longitudinal segmentation, typically achieved through multiple silicon layers interleaved with absorbers. This layered structure enables detailed sampling of the evolving electromagnetic shower, allowing accurate reconstruction of energy deposition, shower development \cite{BARBIELLINI198555}, and particle identification through electron--hadron discrimination.

In the development of large-scale Si--W calorimeters \cite{fabjan2020calorimetry, Bourrion_2023, Sawan_2025}, it is customary to construct small-scale prototypes that retain the essential features of the full detector. Such prototypes are subjected to dedicated test-beam campaigns to validate detector concepts and assess their performance prior to full-scale construction. However, practical constraints arising from the availability or readiness of detector modules, readout electronics, or any other unforeseen situations may restrict the construction of a fully instrumented prototype. Under such circumstances, it is important to investigate whether a reduced-scale prototype can still provide meaningful information about calorimeter performance.

One possible implementation of a resource-constrained prototype is a sequential sampling approach. In this scheme, a single silicon detector layer is placed downstream of a variable number of tungsten absorber depths. By varying the absorber depth across independent measurements, different stages of electromagnetic shower development can be sampled. The resulting measurements can be combined to reconstruct the average longitudinal shower profile. A closely related resource constrained prototype has recently been tested experimentally at CERN PS beamline \cite{Sawan_2026}. However, unlike a fully instrumented calorimeter, such an approach does not preserve event-by-event correlations between different sampling layers and therefore cannot provide direct information on shower fluctuations or intrinsic calorimetric energy resolution. In addition, contributions from particles back-scattered from downstream detector elements are not fully represented.

A recent GEANT4 study of a 20-layer Si--W calorimeter with 3.5 mm tungsten absorbers and 320~$\mu$m silicon sensors investigated the electromagnetic response, longitudinal shower development, linearity, and energy resolution \cite{Potlog:2025}. The present work also employs a similar fully instrumented configuration to provide a reference for comparison with the reduced sequential-sampling configuration. The primary objective of this work is however, to investigate the extent to which such a sequential sampling configuration can reproduce the principal calorimetric characteristics of a fully instrumented Si--W calorimeter. To this end, detailed GEANT4 \cite{AGOSTINELLI2003250} simulations are performed for both a reduced sequential-sampling configuration (Configuration-I) and an ideal fully instrumented calorimeter (Configuration-II). The detector responses are compared in terms of longitudinal shower development, energy deposition, shower maximum position, linearity, and integrated shower observables. Furthermore, the correlation between the responses of the two configurations is studied to establish a response-mapping procedure that relates the reduced configuration to the fully instrumented calorimeter. Since the reduced configuration does not retain event-by-event information across longitudinal sampling layers, its calorimetric energy resolution is not measured directly; instead, an effective resolution is inferred through its response relation to Configuration-II. Through this comparison, we assess the suitability of a resource-constrained sequential-sampling detector for calorimeter characterization and test-beam applications \cite{Muhuri_2020, Sawan_2024, Aehle_2024, MUHURI201424}.

\section{GEANT4 Simulation Framework}
GEANT4 \cite{AGOSTINELLI2003250} is a Monte Carlo toolkit widely used for simulating the particles transport through matter. It provides a detailed description of electromagnetic and hadronic interactions over a wide energy range from eV to GeV, together with flexible geometry construction and event-by-event particle tracking. GEANT4 framework enables the complete modeling of detector response, from primary particle generation to energy deposition in the sensitive detector volumes. For this work GEANT4 version 11.4.0 is used interfaced with ROOT \cite{root} software for the analysis of simulated data. Energy deposited in each silicon detector element was recorded on an event-by-event basis and subsequently analyzed to reconstruct shower profiles and parameters relevant to assess calorimeter performance.

The choice of appropriate Physics list is important for a realistic description of particle transport and energy deposition over the energy range considered in this study. The physics list implemented in the simulation includes electromagnetic, hadronic, decay, and ion interaction processes. Electromagnetic interactions are modeled using the G4EmStandardPhysics package, which incorporates ionization, bremsstrahlung, multiple scattering, Compton scattering, pair production, and photoelectric interactions.

Hadronic interactions are described using G4HadronPhysicsQGSP\_BERT models together with G4HadronElasticPhysicsHP for high-precision treatment of elastic scattering processes. Decays of unstable particles are handled by the G4DecayPhysics package, while interactions involving ions and nuclear fragments are described by G4IonPhysics.

For primary particle generation the G4ParticleGun class used. Electrons with energies of 1, 5, 10, 20, 30, 40, 50, 60, 70, 80, 90, 100, and 120 GeV were simulated to investigate the calorimeter response over a broad energy range. The primary particles were projected perpendicular to the calorimeter surface with their initial position fixed at (x,y)=(0,0) from a longitudinal distance of 20 mm relative to the detector coordinate system. For each incident energy, 20,000 events were generated to ensure adequate statistical precision while maintaining reasonable computational load.
In addition to electrons, 10 GeV positive pions were simulated to study the detector response to minimum-ionizing particles (MIPs) and to illustrate the distinction between electromagnetic and hadronic energy deposition patterns.
\begin{table}[ht]
   \centering
    \begin{tabular}{p{180pt}| p{70pt}}
    \hline
       Material Properties  & Si \\
       \hline
        Atomic Number & 14 \\
        Band Gap (eV) & 1.12 \\
        Density (g/$cm^{3}$) & 2.33 \\
        Mean Energy to create an $e^- h^+$ pair (eV) & 3.62 \\
        Electron mobility ($cm^{2}/Vs$) & 1500 \\
        Carrier life time & $1\mu s$ to 1 ms \\
        Yield & 1.45$\times$$10^{10}$$cm^{-3}$  \\
        \hline
    \end{tabular}
    \caption{Physical properties of silicon material.}
    \label{tab:1}
\end{table}
\FloatBarrier
\section{Detector Design and Geometry}
A world volume of dimensions 50 cm × 50 cm × 50 cm was defined to contain the complete detector assembly. Within this volume, a calorimeter of dimensions 30 cm × 30 cm × 40 cm was constructed. The world volume and calorimeter were filled with the Geant4 predefined material G4\_AIR. 
Silicon was selected as the active detector material owing to its excellent radiation tolerance \cite{rd50}, high charge collection efficiency, and widespread use in modern particle physics experiments. Physical properties of silicon relevant to detector operation are summarized in Table~\ref{tab:1}. Tungsten was selected as the absorber material because of its high density and short radiation length, allowing electromagnetic showers to develop within a compact detector volume.

\begin{figure}[htbp]
    \centering
    \includegraphics[width=0.7
    \linewidth]{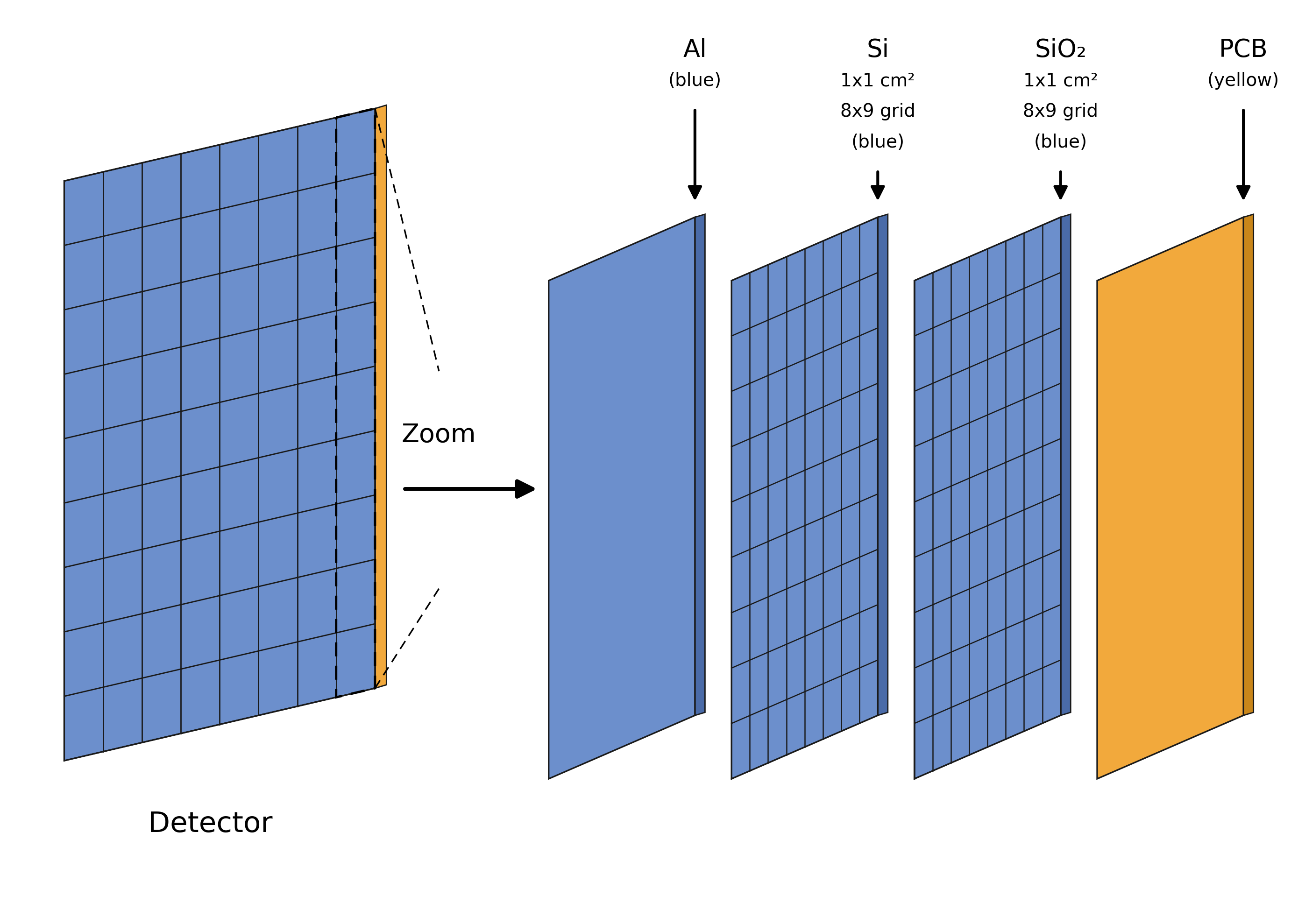}
    \caption{GEANT4 Simulated single unit detector consisting of Al, Si, $\mathrm{SiO_2}$, PCB of thickness 20 $\mu$m, 320 $\mu$m, 0.25 $\mu$m and 2 mm, respectively. For more details refer to text.}
    \label{fig:DetUnit}
\end{figure}
\FloatBarrier
The basic detector module, shown in Fig.~\ref{fig:DetUnit}, consists of four components: a 20 $\mu$m aluminium layer, a 0.25 $\mu$m silicon dioxide layer, a 320 $\mu$m silicon sensor layer, and a 2 mm thick Printed Circuit Board (PCB) of G10 material equivalent [H (6.7\%) + C (27.8\%) + O (45.5\%) + Si (20\%)]. The silicon detector is segmented into an (8 $\times$ 9) array of pads, each with an area of (1$\mathrm{cm^2}$). Individual silicon pads are defined as sensitive detector volumes, enabling event-by-event recording of deposited energy.
Tungsten absorber plates of transverse area 8 $\times$ 9 cm$^{2}$ and thickness 3.5 mm, corresponding approximately to one radiation length, are used. The absorber and detector dimensions have already been optimized to ensure adequate containment of electromagnetic shower over the energy range considered in this study \cite{ALICE_FoCal_TDR_2024}. The corresponding material properties used in the simulation are also summarized in Table~\ref{tab:2}.

\begin{table}[ht]
    \centering
    \begin{tabular}{p{60pt} |p{70pt} |p{70pt}  |p{70pt} |p{70pt}}
    \hline
       Material  & Thickness & Density (g/$cm^{3}$) &Radiation length (cm)  &Molière radius (cm)\\
       \hline
        Si & 320 $\mu$m & 2.33 &9.36  &9.37\\
        $\mathrm{SiO_2}$ & 0.25 $\mu$m & 2.20 &12.01  &9.0\\
        Al & 20 $\mu$m & 2.70 &8.897  &8.9\\
        $PCB_{G10} $ & 2 mm / 4 mm & 1.70 &19.958  &6.53\\
        W & 3.5 mm & 19.3 &0.3504  &0.93\\
        
        \hline
    \end{tabular}
    \caption{Physical properties of different materials used in this simulation.}
    \label{tab:2}
\end{table}
\FloatBarrier

\begin{figure}[htbp]
    \centering
    \includegraphics[width=1.3\linewidth]{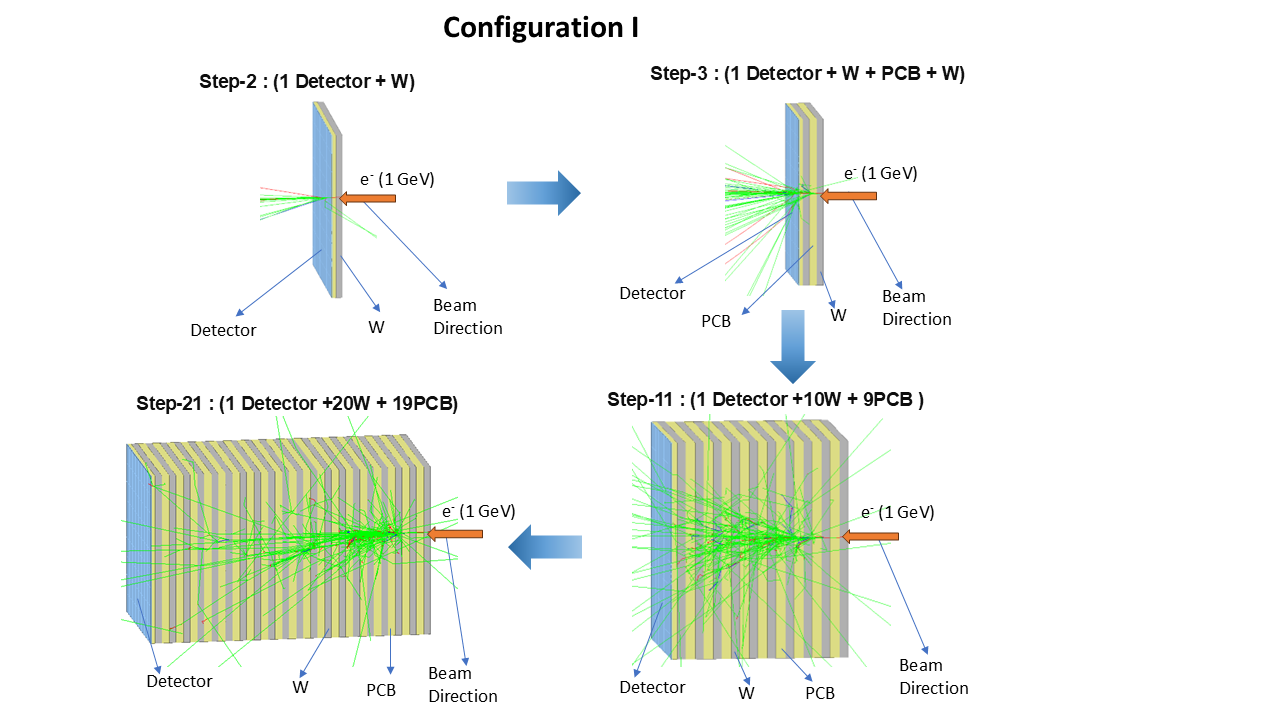}
    \caption{Sequential sampling approach by varying the absorber depth. Shown here is the step 2, 3, 11, 21 and GEANT4 visualization of 1 GeV incident electrons interaction with absorber and detector materials. For more details refer text.}
    \label{fig:ConfigI-setup}
\end{figure}
\FloatBarrier
\subsection{Configuration-I: Sequential Sampling Prototype}
Configuration-I represents a reduced prototype designed to investigate electromagnetic shower development using a sequential sampling strategy. The geometry consists of a single silicon detector layer positioned downstream of progressively increasing absorber thicknesses.

The configuration is implemented through a series of twenty sampling steps. Beginning with a single detector module, additional tungsten absorber and PCB layers are introduced incrementally. Each step therefore corresponds to a different sampling depth within the calorimeter. By combining measurements performed at different absorber depths, the average longitudinal shower profile can be reconstructed. Steps for setting up Configuration-I is pictorially demonstrated in Fig.~\ref{fig:ConfigI-setup}.

This approach significantly reduces the number of silicon detector layers required while retaining the capability to study average shower development throughout the calorimeter volume.

\begin{figure}[htbp]
    \centering
    \includegraphics[width=1.1\linewidth]{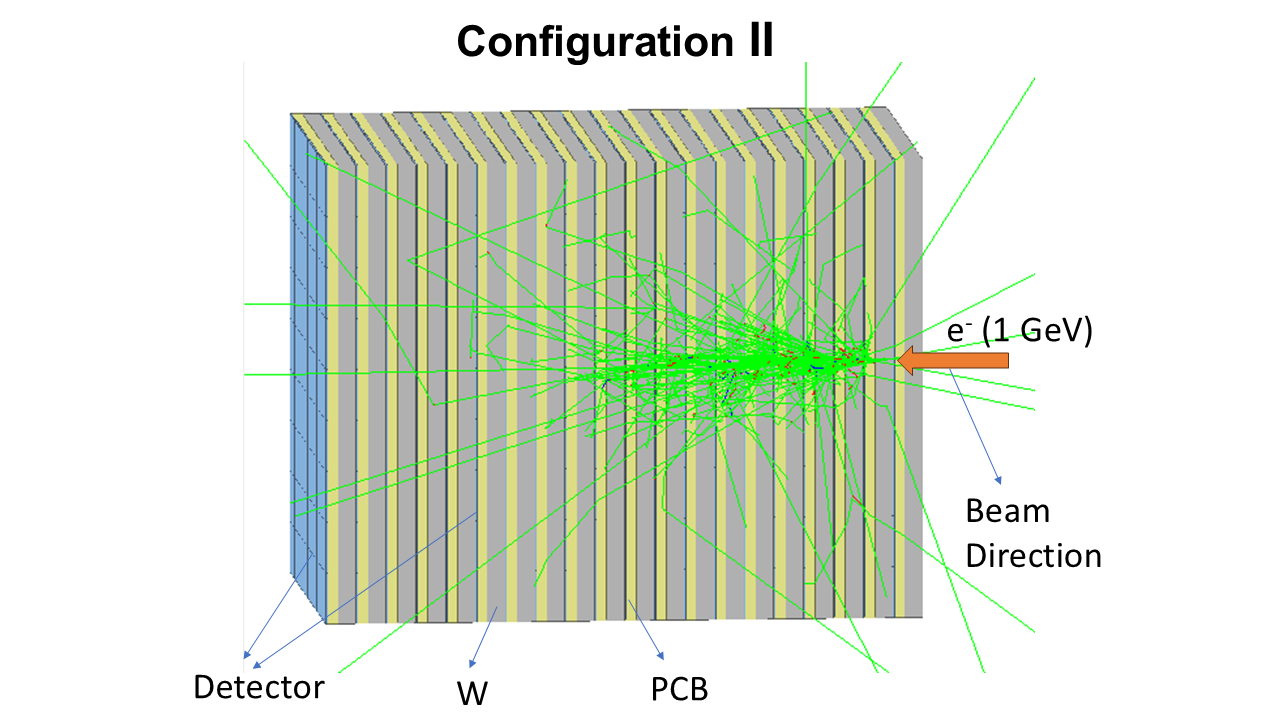}
    \caption{GEANT4 simulated Silicon-tungsten sampling calorimeter consisting of twenty silicon detector layers (individual area of 1 $cm^{2}$ ) interleaved with twenty tungsten absorbers (1 $X_{0}$ thick). GEANT4 visualization of 1 GeV incident electrons interaction for Configuration-II.}
    \label{fig:configII-setup}
\end{figure}
\FloatBarrier
\subsection{Configuration-II: Simultaneous Sampling Prototype}

Configuration-II corresponds to an ideal Si–W sampling calorimeter consisting of twenty silicon detector layers interleaved with twenty tungsten absorbers.

Each sampling layer is composed of a tungsten absorber followed by a detector module consisting of aluminum, silicon dioxide, silicon, and PCB materials. The resulting geometry, as shown in Fig~\ref{fig:configII-setup}, provides continuous longitudinal sampling of the electromagnetic shower throughout the calorimeter depth.

Unlike Configuration-I, all sampling layers are instrumented simultaneously, enabling event-by-event reconstruction of shower development and direct measurement of total deposited energy. Configuration-II therefore serves as the reference detector against which the performance of the reduced prototype is evaluated.


\section{Results and Discussion}
\begin{figure}[htbp]
    \centering
    \includegraphics[width=0.7\linewidth]{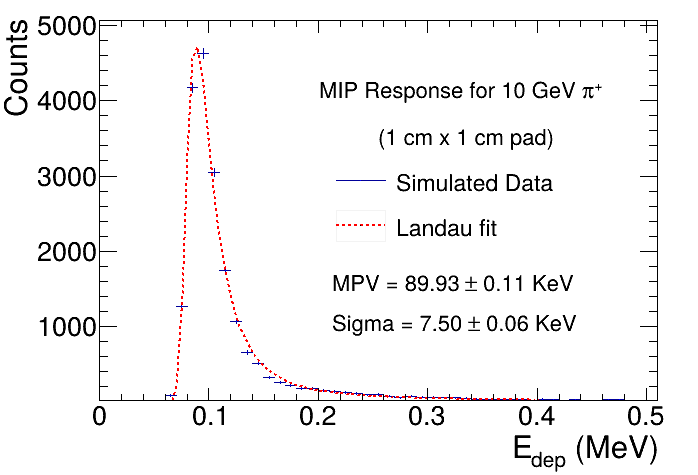}
    \caption{Simulated energy-deposition spectrum for 10 GeV $\pi^{+}$ in a single silicon pad of area 1 $\times 1~\mathrm{cm}^{2}$. The distribution exhibits the characteristic response of a minimum-ionizing particle (MIP), with a most probable energy deposition of approximately 89 KeV.}
    \label{fig:MIP-pi}
\end{figure}

As a first step in evaluating the performance of the calorimeter, the energy deposition in a single silicon pad of area (1~$\mathrm{cm}^2$) has been investigated using a 10 GeV ($\pi^+$) beam. At this energy, charged pions predominantly behave as minimum-ionizing particles (MIPs), producing a characteristic energy deposition with a most probable value (MPV) of approximately 89 keV in the silicon sensor, as shown in Fig.~\ref{fig:MIP-pi}.
The MIP response provides a basic sanity check of the simulated silicon energy-deposition response and establishes the characteristic signal scale for a minimum-ionizing particle.

The response of the Si--W calorimeter has been investigated using two detector configurations described in previously- a reduced sequential-sampling prototype (Configuration-I) and a fully instrumented sampling calorimeter (Configuration-II). Although both configurations are exposed to identical incident particles, their sampling strategies differ substantially. Configuration-I reconstructs the average longitudinal shower profile by performing measurements at different absorber depths using a single active detector layer, whereas Configuration-II records the complete shower development simultaneously in all sampling layers on an event-by-event basis.
In the following sections, the calorimetric performance of the two configurations is evaluated and compared in terms of longitudinal shower development, shower maximum position, calorimetric response, linearity, energy containment, and energy resolution.
\FloatBarrier
\subsection{Configuration-I}

\begin{figure}[htbp]
    \centering
    \includegraphics[width=1.0\linewidth]{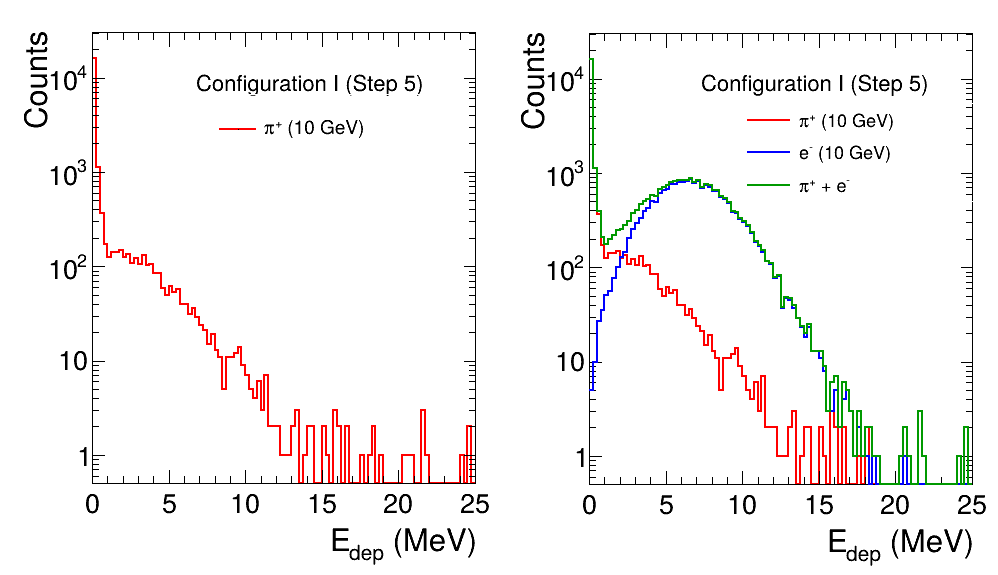}
    \caption{Energy-deposition spectra measured in the fourth sampling layer of Configuration-I for 10 GeV incident particles. (Left) Response to a 10 GeV pion beam, exhibiting the characteristic minimum-ionizing particle (MIP) signal. (Right) Comparison of the energy-deposition spectra for 10 GeV electrons, 10 GeV pions, and their combined response, demonstrating the clear separation between EM showers and MIP signals.}
    \label{fig:Edep-pi-e-10GeV}
\end{figure}

Figure~\ref{fig:Edep-pi-e-10GeV} presents the energy deposited in the fourth sampling layer of Configuration-I for pion and electron beams of energy 10 GeV. The response to 10 GeV pion beams, shown in the left panel, exhibits a narrow energy-deposition distribution, a characteristic of MIP. In contrast, the response to 10 GeV electrons, shown in the right panel, is significantly broader and shifted towards higher deposited energies. This behavior arises from the development of electromagnetic(EM) showers in the tungsten absorbers through bremsstrahlung and pair-production processes.

Also shown in the right panel of Fig~\ref{fig:Edep-pi-e-10GeV} is the combined response of 10 GeV electron and pion beams. The resulting distribution reflects contributions from both MIPs and EM showers, exhibiting features associated with each component. The clear separation between the pion and electron responses demonstrates the capability of the Si--W calorimeter in Configuration-I to distinguish EM showers from MIPs. Such discrimination is an important requirement for effective particle identification using sampling calorimeters in contemporary high-energy physics experiments.

\begin{figure}[htbp]
    \centering
    \includegraphics[width=0.8\linewidth]{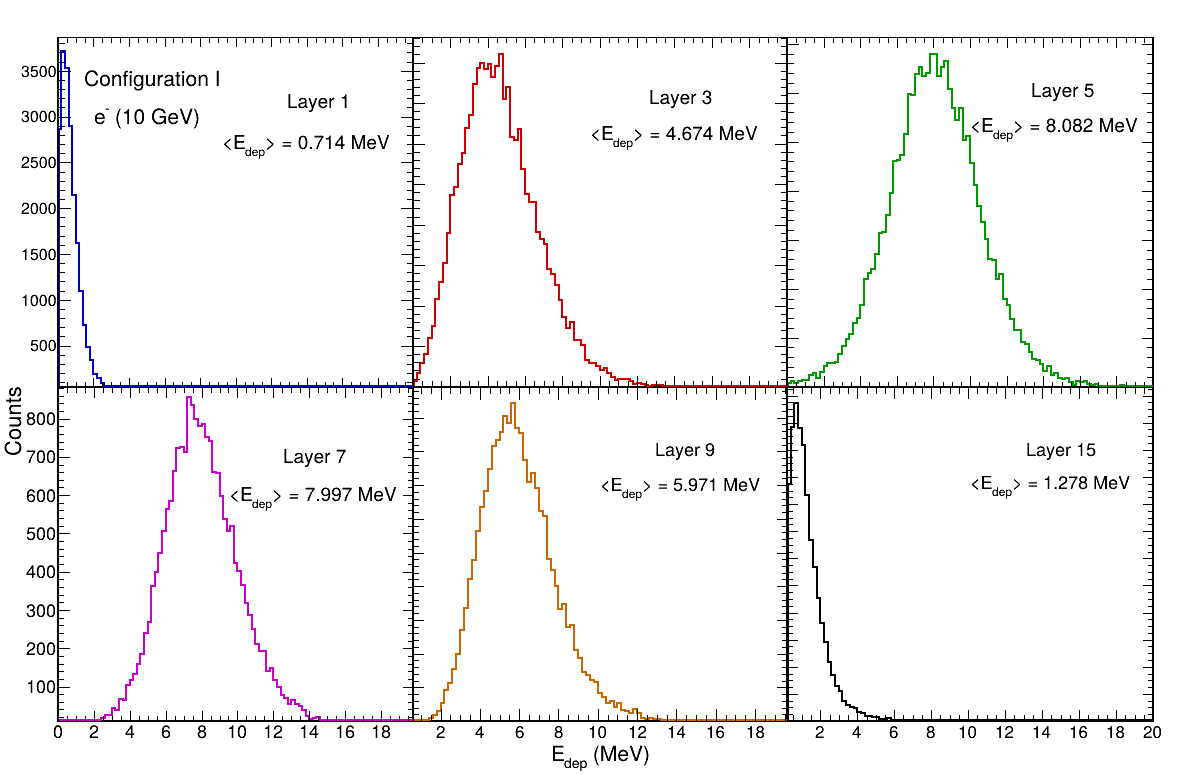}
    \caption{Energy-deposition spectra for 10 GeV incident electrons recorded in the detector layers 1, 3, 5, 7, 9, and 15 of the Si--W calorimeter in Configuration-I. }
    \label{fig:Edep-e-diff-layers}
\end{figure}

A representative energy-deposition spectra obtained at different sampling depths are shown in Fig.~\ref{fig:Edep-e-diff-layers}. As the EM shower develops within the absorber material, the average deposited energy initially increases with depth owing to the multiplication of secondary particles. The energy deposition reaches a maximum near the shower core and subsequently decreases as the shower attenuates. The observed evolution of the energy-deposition spectra provides a direct indication of the longitudinal EM shower development in the calorimeter.

\FloatBarrier

\begin{figure}[htbp]
    \centering
    \includegraphics[width=0.7\linewidth]{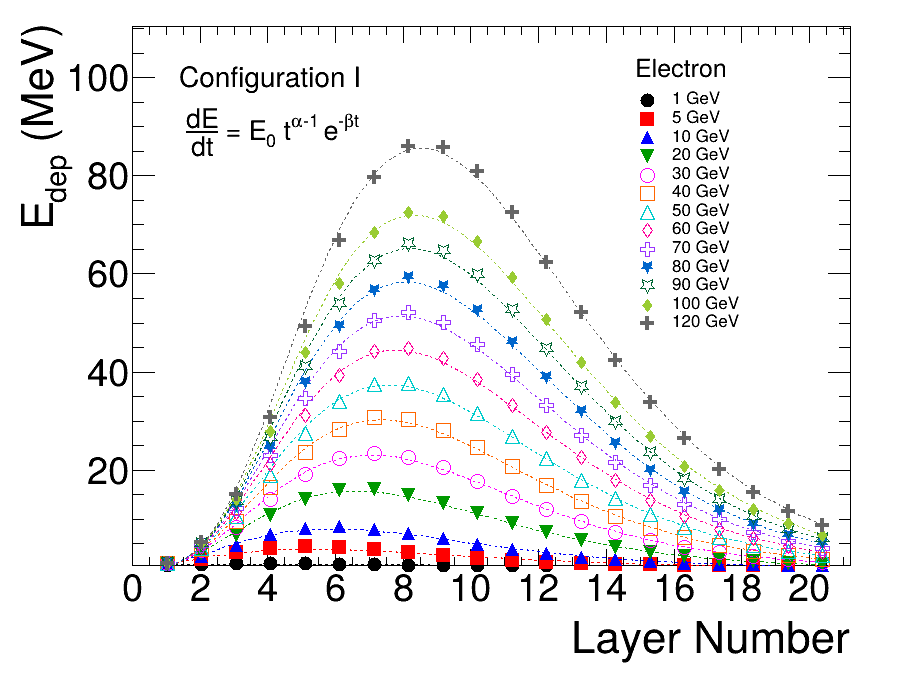}
    \caption{Longitudinal shower profiles of the Si--W calorimeter for incident electron energies ranging from 1 to 120 GeV as a function of calorimeter depth for Configuration-I. The profiles are fitted with the empirical gamma-distribution parameterization given by Eq.~(\ref{eq:gamma function}).}
    \label{fig:2}
\end{figure}

\FloatBarrier
Eventually, the longitudinal development of EM showers was studied by recording the mean energy deposited in each silicon sampling layer as a function of calorimeter depth. Figure~\ref{fig:2} presents the resulting longitudinal shower profiles for incident electron energies ranging from 1 to 120 GeV. In the present detector geometry, each sampling layer corresponds to approximately 1.012 X$_{0}$.
\FloatBarrier
For all incident energies, the deposited energy initially increases with depth as the number of secondary particles produced in EM cascade grows. The energy deposition reaches a maximum at a characteristic depth, commonly referred to as the shower maximum, and subsequently decreases as the shower attenuates.

The longitudinal shower profiles are parameterized using the gamma-distribution form \cite{LONGO1975283} commonly employed to describe electromagnetic shower development,

\begin{equation}
\frac{dE}{dt}
=
\frac{E_{0}\beta(\beta t)^{(\alpha-1)} e^{-\beta t}}
{\Gamma(\alpha)} \approx E_{0}t^{\alpha-1}e^{-\beta t}
\label{eq:gamma function}
\end{equation}

where (t) denotes the calorimeter depth measured in units of radiation length, (E$_{0}$) is a normalization parameter, and ($\alpha$) and ($\beta$) characterize the shape of the shower profile. The parameterization provides a good description of the simulated longitudinal distributions over the entire energy range investigated.

As the incident electron energy increases, the shower maximum shifts towards larger depths, reflecting the longer development length of higher-energy electromagnetic cascades. The observed behavior is consistent with the expected of EM shower development and demonstrates that the detector geometry implemented in the reduced prototype in the simulation reproduces the characteristic features of shower development.
\begin{figure}
    \centering
    \includegraphics[width=0.7\linewidth]{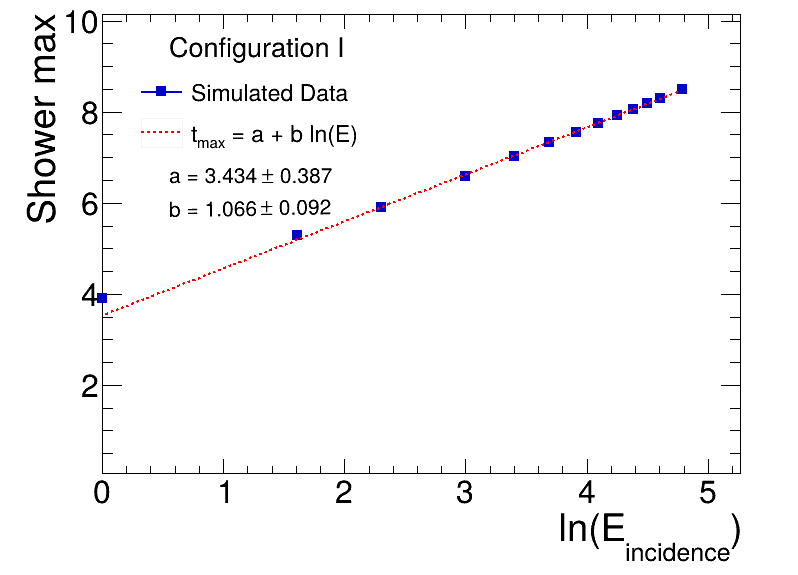}
    \caption{Position of the shower maximum ($t_{\mathrm{max}}$) in the Si--W calorimeter as a function of the incident electron energy for Configuration-I. }
    \label{fig:4}
\end{figure}
\FloatBarrier

The position of the shower maximum was then extracted from the fitted longitudinal shower profiles for each incident electron energy in terms of fit parameters, a and b, as $t_{max}$ = $\frac{\alpha-1}{\beta}$. Figure~\ref{fig:4} shows the dependence of the shower maximum position on the incident electron energy.
As expected, the depth of the shower maximum increases approximately linearly with the logarithm of the incident electron energy. Higher-energy particles produce a larger number of secondary particles and therefore require a greater absorber thickness before the shower reaches its maximum intensity.
As a result shower maximum moves progressively deeper into the calorimeter with increasing beam energy. 

The determination of the shower maximum is of practical importance in calorimeter design because it provides guidance on the detector depth required for efficient energy containment. In addition, knowledge of the shower maximum location can be used to optimize the placement of highly granular detector layers \cite{de_Haas_2018}
in order to enhance particle identification and improve energy reconstruction performance. The results presented here therefore provide useful information for the design and optimization of Si--W calorimeters operating over a broad energy range.

\begin{figure}[htbp]
    \centering
    \includegraphics[width=0.7\linewidth]{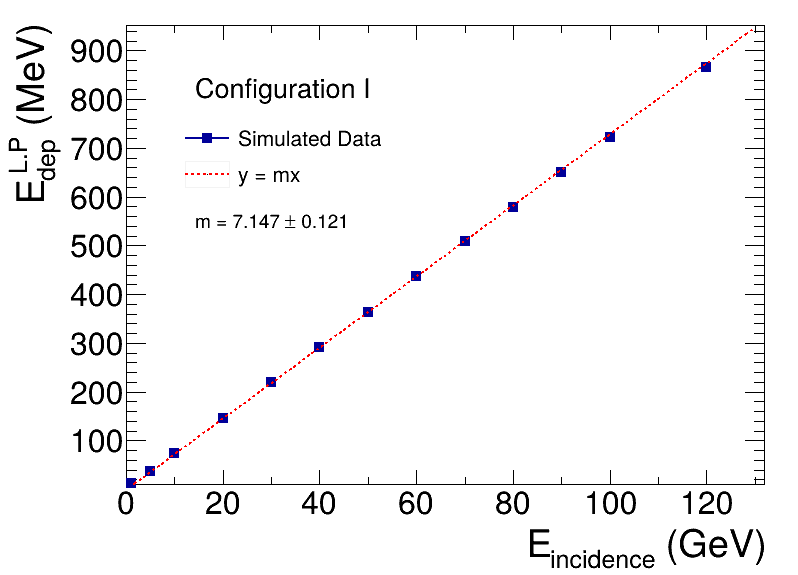}
    \caption{Reconstructed calorimetric response as a function of the incident electron energy for Configuration-I. The reconstructed energy (E$_{\mathrm{dep}}^{\mathrm{L.P}}$) is obtained by integrating the fitted longitudinal shower profile over the physical depth of the calorimeter (20 sampling layers). }
    \label{fig:3}
\end{figure}

For Configuration-I, the total calorimetric response is now obtained by integrating the fitted longitudinal shower profile over the full detector depth. The resulting reconstructed energy is shown as a function of the incident electron energy in Figure~\ref{fig:3}.
\FloatBarrier
A linear dependence of the reconstructed energy on the incident beam energy is observed throughout the investigated energy range. The response is well described by a first-order polynomial, indicating that the calorimeter preserves linearity over the full range of simulated energies. Such behavior is a fundamental requirement for calorimetry, as it ensures a direct correspondence between the measured detector signal and the energy of the incident particle.
The observed linear response further demonstrates that the longitudinal shower profile reconstructed using the sequential-sampling approach provides a reliable estimate of the overall calorimetric response.

However, the finite depth of the calorimeter entails an incomplete containment of EM showers. To investigate this effect, the fraction of shower energy escaping beyond the final sampling layer was evaluated as a function of incident electron energy.
The longitudinal shower profiles give a hint that the leakage contribution increases with beam energy because the shower maximum moves deeper into the detector and a larger fraction of secondary particles reach the end of the calorimeter volume. To quantify this effect, the longitudinal leakage was estimated by comparing the total energy deposited over an infinite calorimeter depth, $E_{\mathrm{dep}}^{\infty}$, obtained from the fitted gamma-distribution using eq.(\ref{eq:EdepInf}), with the energy deposited within the physical depth of the calorimeter, i.e, $E_{\mathrm{dep}}^{20}$ calculated using eq.(\ref{eq:Edep20X0}) . The longitudinal leakage energy and the corresponding leakage fraction were calculated using equations \ref{eq:leakage} and \ref{eq:leakage_percentage}, respectively. This procedure provides an estimate of the fraction of the EM shower that escapes the calorimeter due to its finite longitudinal extent.
\begin{figure}[htbp]
    \centering
    \includegraphics[width=0.9\linewidth]{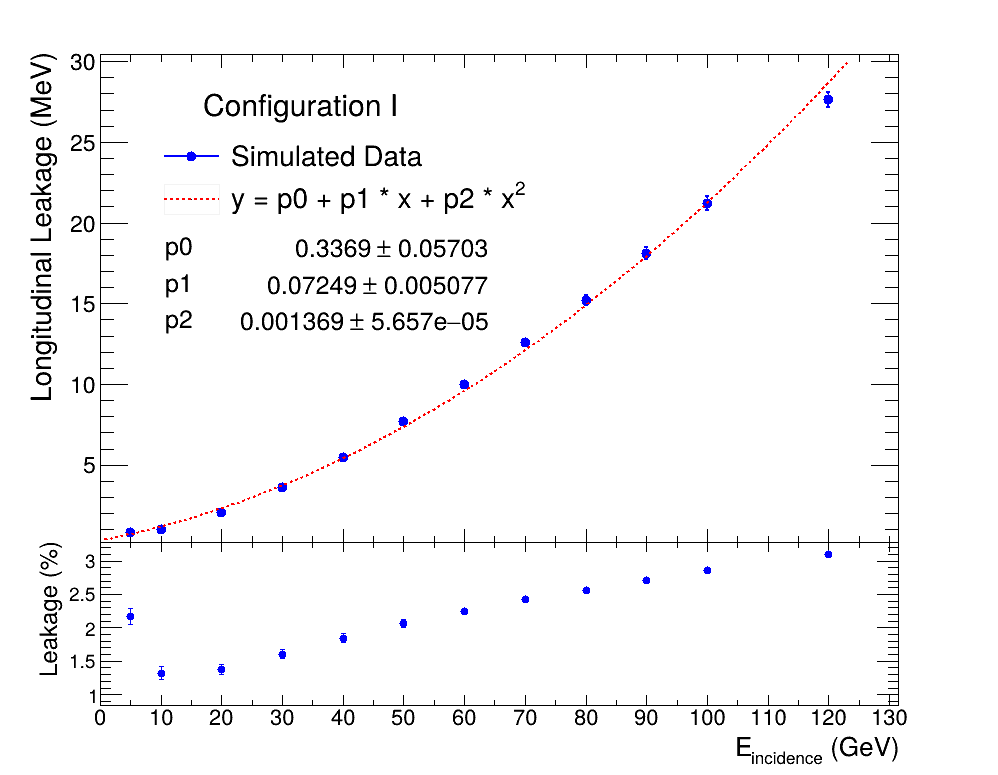}
    \caption{(Top)Estimated longitudinal energy leakage and (Bottom) leakage percentage, as a function of incident electron energy for Configuration-I.}
    \label{fig:500}
\end{figure}
\FloatBarrier

\begin{equation}
E_{\mathrm{dep}}^{\infty}
=
\int_{1.018}^{\infty}
E_{0}\, t^{\alpha} e^{-\beta t}\, dt
\label{eq:EdepInf}
\end{equation}

\begin{equation}
E_{\mathrm{dep}}^{20}
=
\int_{1.018}^{20.393}
E_{0}\, t^{\alpha} e^{-\beta t}\, dt
\label{eq:Edep20X0}
\end{equation}

\begin{equation}
\text{Longitudinal Leakage (L.L.)}
=
E_{\mathrm{dep}}^{\infty}
-
E_{\mathrm{dep}}^{20}
\label{eq:leakage}
\end{equation}

The leakage percentage is-

\begin{equation}
\text{Leakage Percentage}
=\frac{\mathrm{E_{\mathrm{dep}}^{\infty}
-
E_{\mathrm{dep}}^{20}}}
     {E_{\mathrm{dep}}^{\infty}}
\times 100
\label{eq:leakage_percentage}
\end{equation}

Figure~\ref{fig:500} presents the estimated leakage fraction and leakage percentage  for the studied energy range. The effect is negligible at low energies but becomes increasingly important at higher energies, where the available calorimeter depth approaches the limit of shower containment length. While at lower energies longitudinal leakage is in between 1-2\% level, it grows upto 3\% for the highest incident energy studied here. This observation highlights the importance of balancing detector compactness against energy containment in the design of electromagnetic calorimeters.

\subsection{Configuration-II}
\begin{figure}[htbp]
    \centering
    \includegraphics[width=1.0\linewidth]{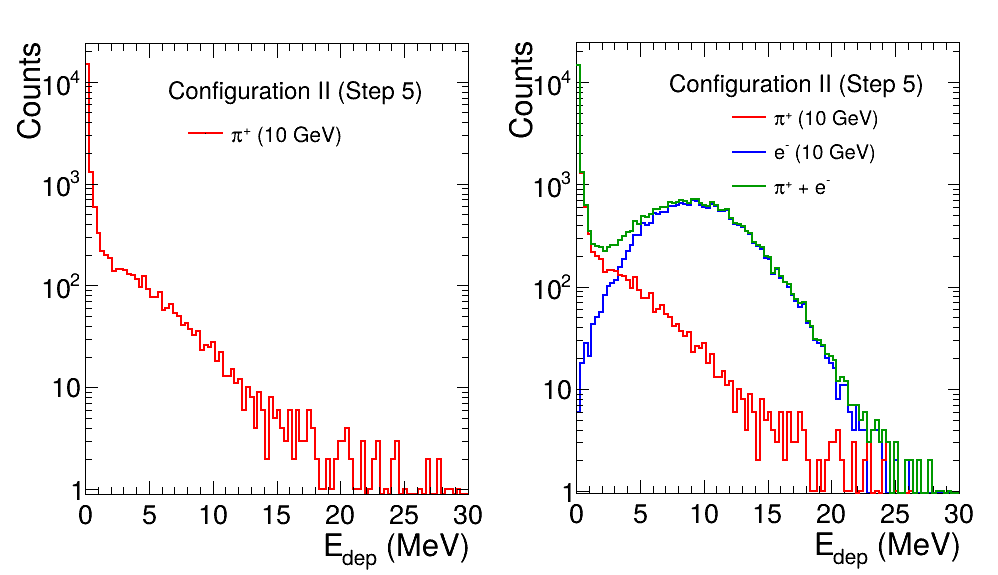}
    \caption{For Configuration-II, the left panel shows the simulated response of silicon detectors to 10 GeV pions for the $4^{th}$ layer and the right panel shows the response of 10 GeV pion and electron for the $4^{th}$ layer.}
    \label{fig:Edep-pi-e-10GeV-configII}
\end{figure}
\begin{figure}[htbp]
    \centering
    \includegraphics[width=1.02\linewidth]{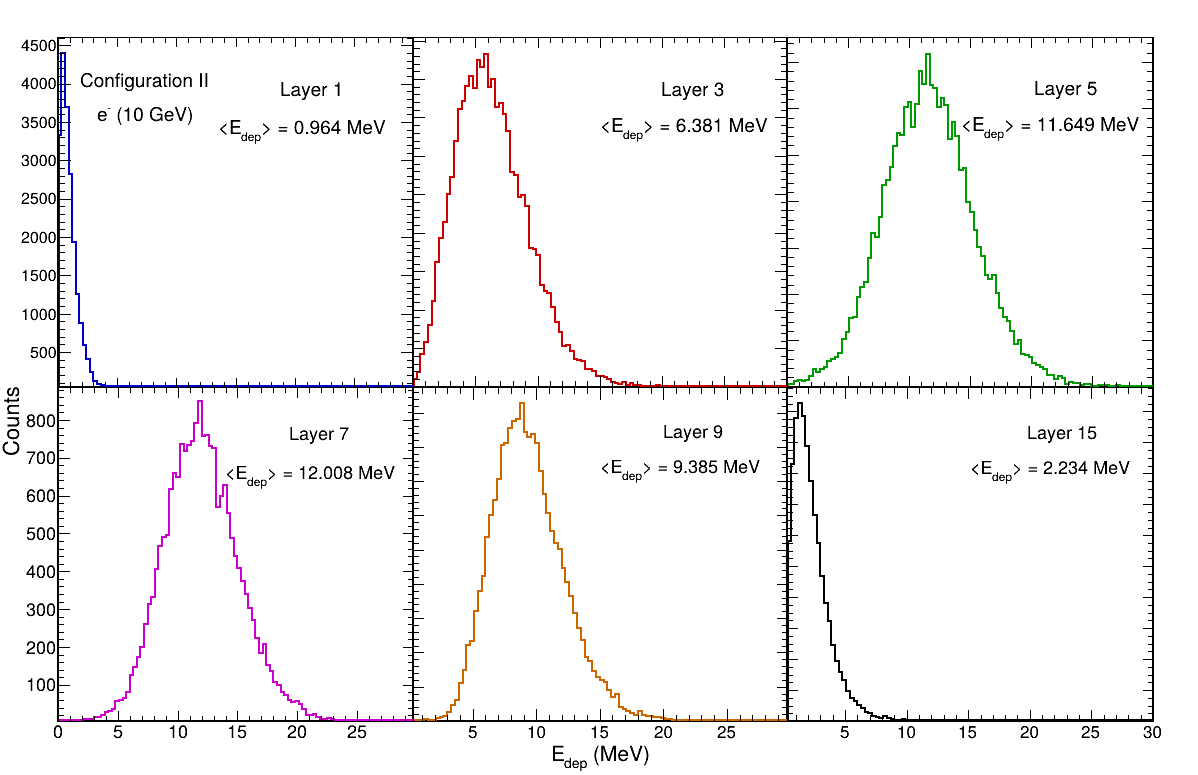}
    \caption{Energy-deposition spectra for 10 GeV incident electrons recorded in the detector layers 1, 3, 5, 7, 9, and 15 of the Si--W calorimeter in Configuration-II.}
    \label{fig:Edep-e-diff-layers-configII}
\end{figure}

\FloatBarrier
We now proceed to discuss results obtained using Configuration-II, which represents the fully instrumented Si-W calorimeter with interleaved tungsten absorbers and silicon detector layers. As in the previous section, the detector response was first investigated by examining the energy deposited in a silicon pad located in the fourth sampling layer for incident 10 GeV electrons and pions.
The corresponding energy-deposition spectra are shown in Fig.~\ref{fig:Edep-pi-e-10GeV-configII}. Also the energy-deposition spectra obtained at different sampling depths of Configuration-II are shown in Fig.~\ref{fig:Edep-e-diff-layers-configII}. The overall characteristics of the distributions are similar to those observed for Configuration-I. 
\FloatBarrier

Although the general features remain unchanged, the magnitude of the energy deposited in each sampling layer is observed to be larger in Configuration-II than in Configuration-I. This difference arises from the difference in detector geometries and sampling strategies employed in the two Configurations. In particular secondary particles originating from neighboring layers and particles back-scattered from downstream detector elements can enhance the energy deposition recorded in a given silicon layer.

The observed increase in deposited energy therefore reflects the more realistic shower development in the fully instrumented calorimeter. Nevertheless, the qualitative similarity of the electron and pion responses in the two Configurations indicates that the reduced prototype captures the essential features of particle interactions in the detector.

\begin{figure}[htbp]
    \centering
    \includegraphics[width=1.0\linewidth]{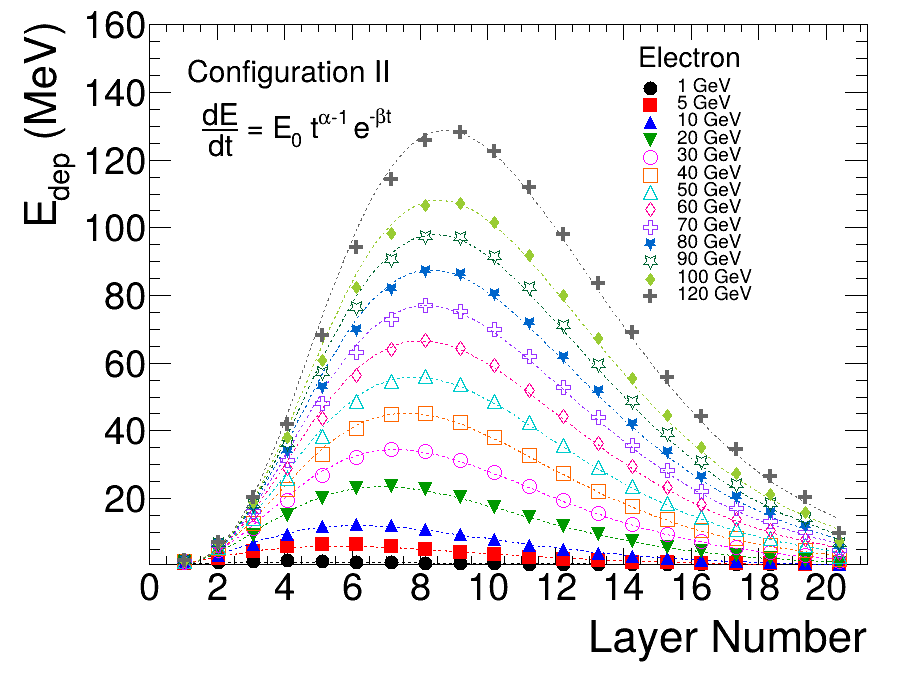}
    \caption{Longitudinal shower profiles of the Si--W calorimeter for incident electron energies ranging from 1 to 120 GeV as a function of calorimeter depth for Configuration-II. The profiles are fitted with the empirical gamma-distribution parameterization given by Eq.~(\ref{eq:gamma function}).}
    \label{fig:6}
\end{figure}
\FloatBarrier
The longitudinal shower profiles obtained for Configuration-II are shown in Fig.~\ref{fig:6}. The overall behavior is found to be qualitatively similar to that observed for Configuration-I. For all incident electron energies, the deposited energy increases with calorimeter depth, reaches a maximum, and subsequently decreases as the shower attenuates within the absorber material.

\begin{figure}[htbp]
    \centering
    \includegraphics[width=0.6\linewidth]{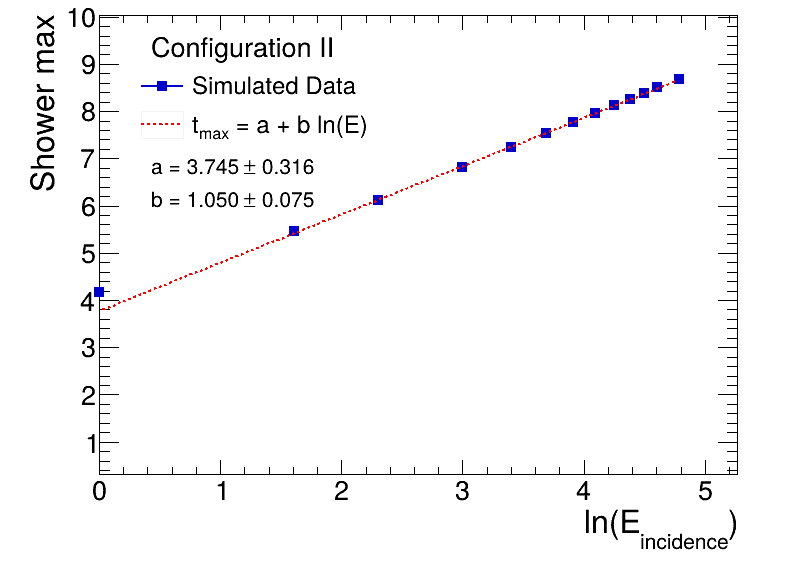}
    \caption{Position of the shower maximum ($t_{\mathrm{max}}$) in the Si--W calorimeter as a function of the incident electron energy for Configuration-II.}
    \label{fig:8}
\end{figure}

The profiles are well described by the gamma-distribution parameterization introduced in equation~\ref{eq:gamma function}. As expected, the shower maximum shifts towards larger depths with increasing incident energy, reflecting the longer development length of higher-energy EM cascades. The extracted shower maximum positions are shown in Fig.~\ref{fig:8} and exhibit the same logarithmic dependence on incident energy observed for Configuration-I.
\FloatBarrier
The similarity of the longitudinal shower profiles and shower maximum positions obtained from the two detector Configurations indicates that the sequential-sampling approach for Configuration-I reproduces the principal features of the average EM shower development. Therefore, in the following discussion we focus on observables that are uniquely accessible in the fully instrumented calorimeter.

\begin{figure}[htbp]
    \centering
    \includegraphics[width=1.0\linewidth]{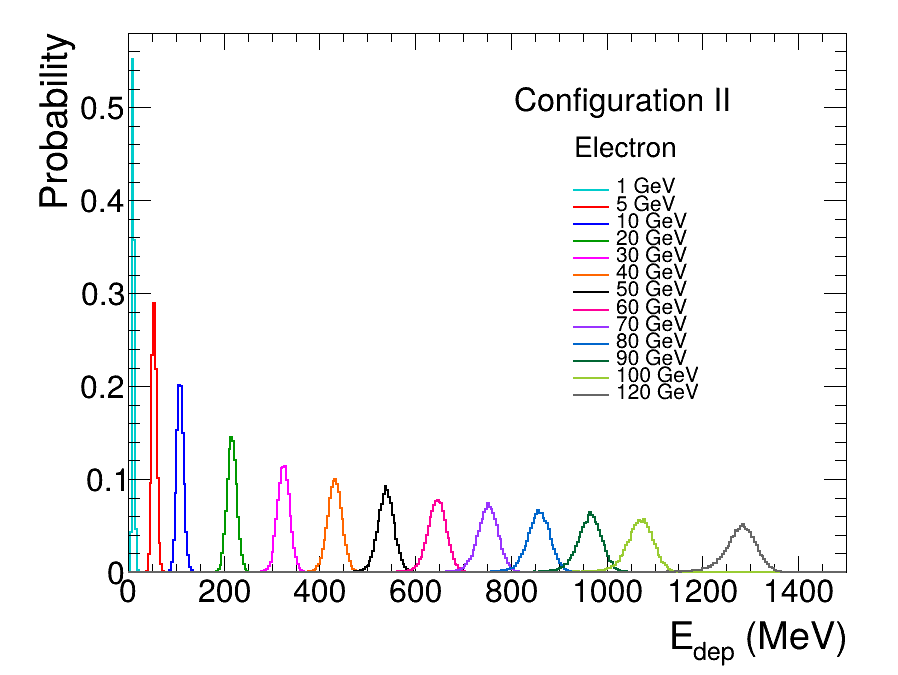}
    \caption{Event-by-event reconstructed energy distributions for incident electrons of energies between 1 and 120 GeV in Configuration-II. The total reconstructed energy is obtained by summing the energy deposited in all silicon pads of the calorimeter.}
    \label{fig:9}
\end{figure}

\begin{figure}[htbp]
    \includegraphics[width=0.50\linewidth]{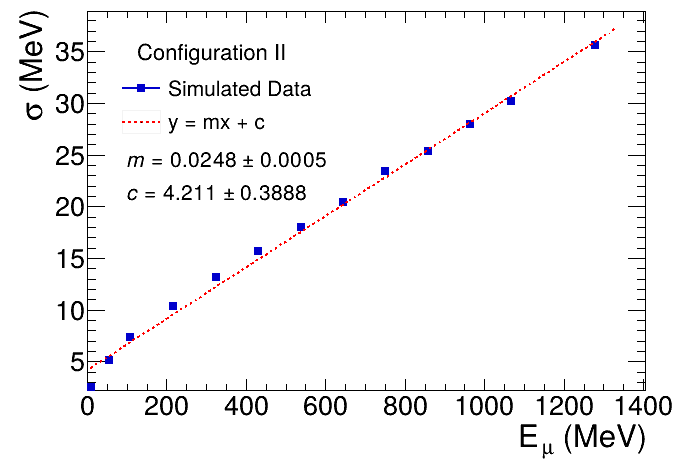}
    \includegraphics[width=0.50\linewidth]{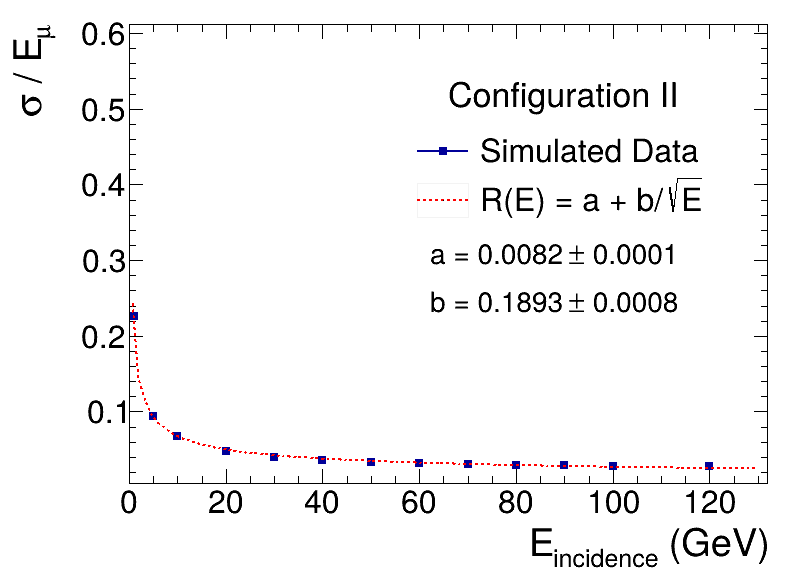}
    \caption{Performance parameters for Configuration-II. (Left) Dependence of the Gaussian width ($\sigma$) on the mean reconstructed energy (E$_{\mu}$), together with a linear fit to the simulated data. (Right) Corresponding energy resolution as a function of the incident electron energy. The simulate-data are well described by the parameterization in Eq.~\ref{eq:resolution-II} with a = 0.0082 $\pm$ 0.0001 and b = 0.1893 $\pm$ 0.0008.}
    \label{fig:11}
\end{figure}

Unlike Configuration-I, where the calorimetric response is reconstructed from the average longitudinal shower profile, Configuration-II enables direct event-by-event measurement of the total deposited energy. Since all twenty silicon sampling layers are instrumented simultaneously, the energy deposited in every silicon pad can be recorded for each event.
\begin{figure}[htbp]
    \centering
    \includegraphics[width=0.55\linewidth]{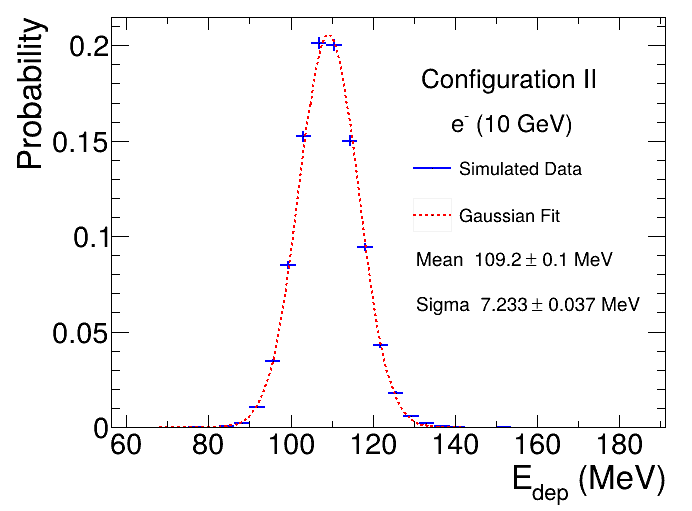}
    \caption{Event-by-event reconstructed energy distribution for 10 GeV incident electrons in Configuration-II. The total reconstructed energy is obtained by summing the energy deposited in all silicon pads of the calorimeter. The distribution is fitted with a Gaussian function to extract the mean reconstructed energy ($\mu$) and the corresponding standard deviation ($\sigma$).}

    \label{fig:202}
\end{figure}
\FloatBarrier
The total reconstructed energy was obtained by summing the energy deposited in all silicon pads of the calorimeter, corresponds to the total energy deposited in all 72$\times$20=1440 silicon pads of the calorimeter.
Figure~\ref{fig:202} shows the event-by-event reconstructed energy distribution for incident electrons of energy 10 GeV. The distribution exhibits a well-defined peak corresponding to the average detector response together with a finite width arising from fluctuations in electromagnetic shower development and the sampling nature of the calorimeter.
The event-by-event reconstructed energy distributions obtained for all investigated beam energies are shown in Fig.~\ref{fig:9}. As the incident energy increases, the distributions shift towards larger reconstructed energies, reflecting the increased energy deposited by higher-energy electromagnetic showers. At the same time, the absolute width of the distributions also increases due to larger shower-to-shower fluctuations in deposited energy.
\FloatBarrier
Each distribution was fitted with a Gaussian function to extract the mean reconstructed energy ($\mu$) and the corresponding standard deviation ($\sigma$). The mean value characterizes the average calorimeter response, while the standard deviation provides a quantification of the event-by-event fluctuations in the reconstructed energy, shown in the left panel of Fig.~\ref{fig:11}. The ratio $\sigma$/$\mu$, which represents the relative fluctuation of the reconstructed energy decreases with increasing incident energy, as shown in the right panel of Fig.~\ref{fig:11}, indicating an improvement in the calorimeter energy resolution at higher energies, consistent with the expected statistical nature of EM shower development.

The energy resolution obtained is well described by the parameterization \cite{ROUENE2013470, RevModPhys.75.1243}
\begin{equation}
    \frac{\sigma}{\mu} = a \oplus \frac{b}{\sqrt{E_{\text{inc}}}},
    \label{eq:resolution-II}
\end{equation}
where "a" represents the constant term and "b" denotes the stochastic term with fitted values of a=0.0082±0.0001 and b=0.1893±0.0008.

\FloatBarrier

A unique feature of Configuration-II is that the total calorimetric response can be determined using two independent methods: (i) integration of the fitted longitudinal shower profile and (ii) direct event-by-event summation of the energy deposited in all silicon pads of the detector. The former approach provides an estimate based on the average shower development, whereas the latter exploits the complete information available from the fully instrumented calorimeter.
\begin{figure}[htbp]
    \includegraphics[width=0.5\linewidth]{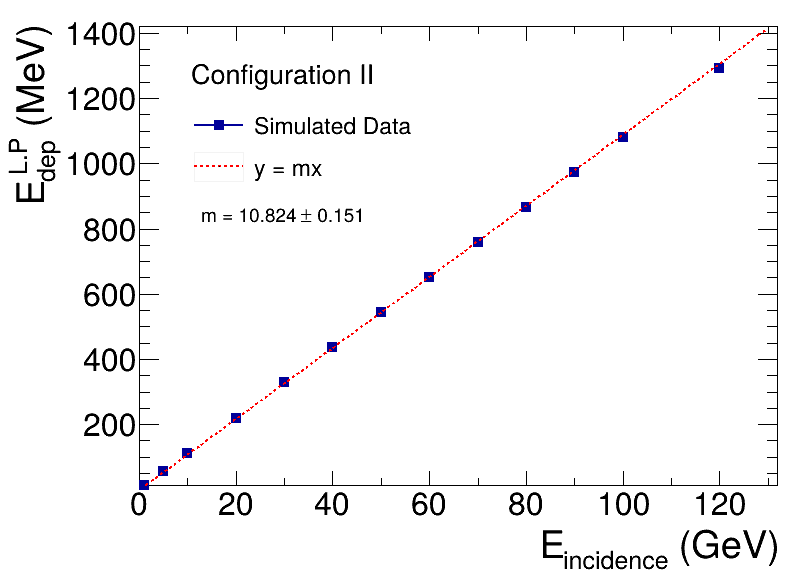}
    \includegraphics[width=0.52\linewidth]{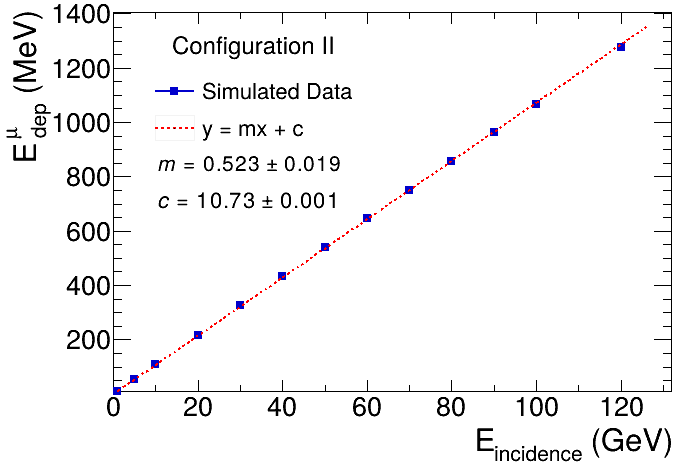}
    \caption{ For Configuration-II, (Left) Reconstructed energy obtained by integrating the longitudinal shower profile over the full calorimeter depth (20 sampling layers) as a function of the incident electron energy. (Right) Mean reconstructed energy (E$_{\mu}$) obtained from Gaussian fits to the event-by-event energy distributions. In both cases, the simulated data are well described by linear fits, demonstrating the linearity of the calorimeter response.}
    \label{fig:7}
\end{figure}
\FloatBarrier

\begin{figure}[htbp]
    \includegraphics[width=0.53\linewidth]{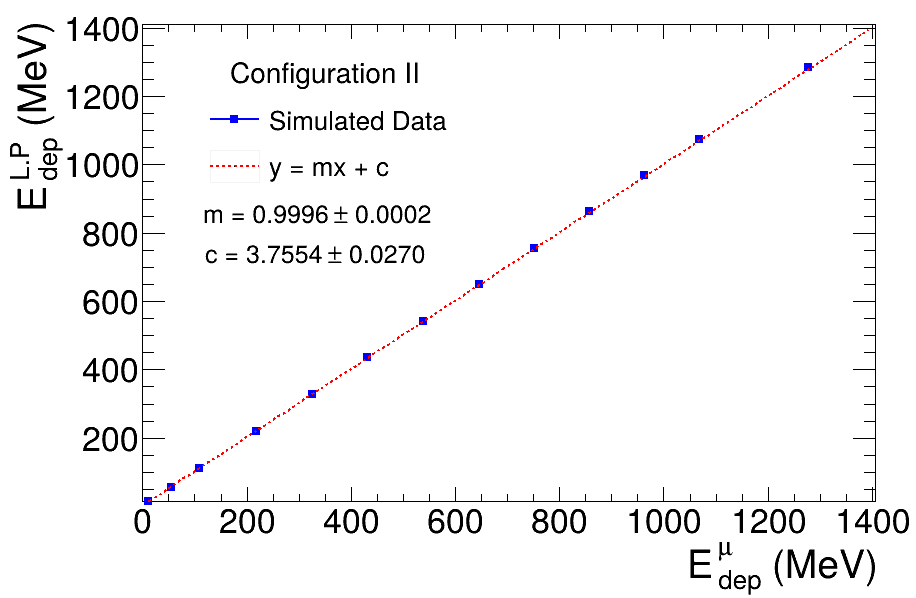}
    \includegraphics[width=0.47\linewidth]{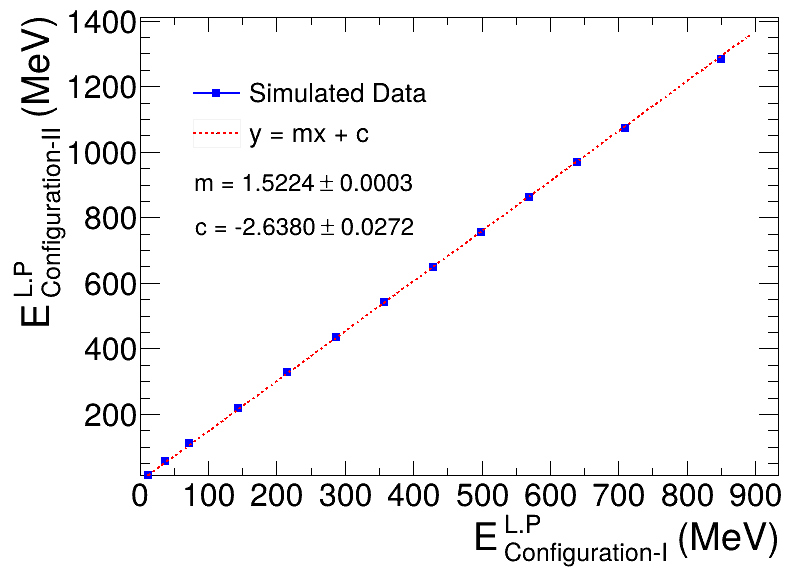}
    \caption{(Left) Correlation between the reconstructed energy obtained from the integrated longitudinal shower profile and the mean reconstructed energy (E$_{\mu}$) extracted from Gaussian fits to the event-by-event energy distributions for Configuration-II. (Right) Correlation between the reconstructed energies obtained from the integrated longitudinal shower profiles of Configurations I and II. In both panels, the simulated data are well described by linear fits. These calibration curves form the basis for mapping the calorimetric response of Configuration-I onto that of the  Configuration-II.}

    \label{fig:13}
\end{figure}
\FloatBarrier

Figures~\ref{fig:7} present the reconstructed calorimetric response obtained using two independent approaches in Configuration-II. The left panel shows the reconstructed energy obtained by integrating the longitudinal shower profile, whereas the right panel presents the mean reconstructed energy extracted from Gaussian fits to the event-by-event energy distributions. Both observables exhibit linear dependence on the incident electron energy over the investigated range. In addition, the reconstructed energies obtained from the two methods are themselves linearly correlated, as shown in the left panel of Fig.~\ref{fig:13}, demonstrating the consistency between the longitudinal-profile and event-by-event reconstruction techniques.

\begin{figure}[htbp]
    \centering
    \includegraphics[width=1.0\linewidth]{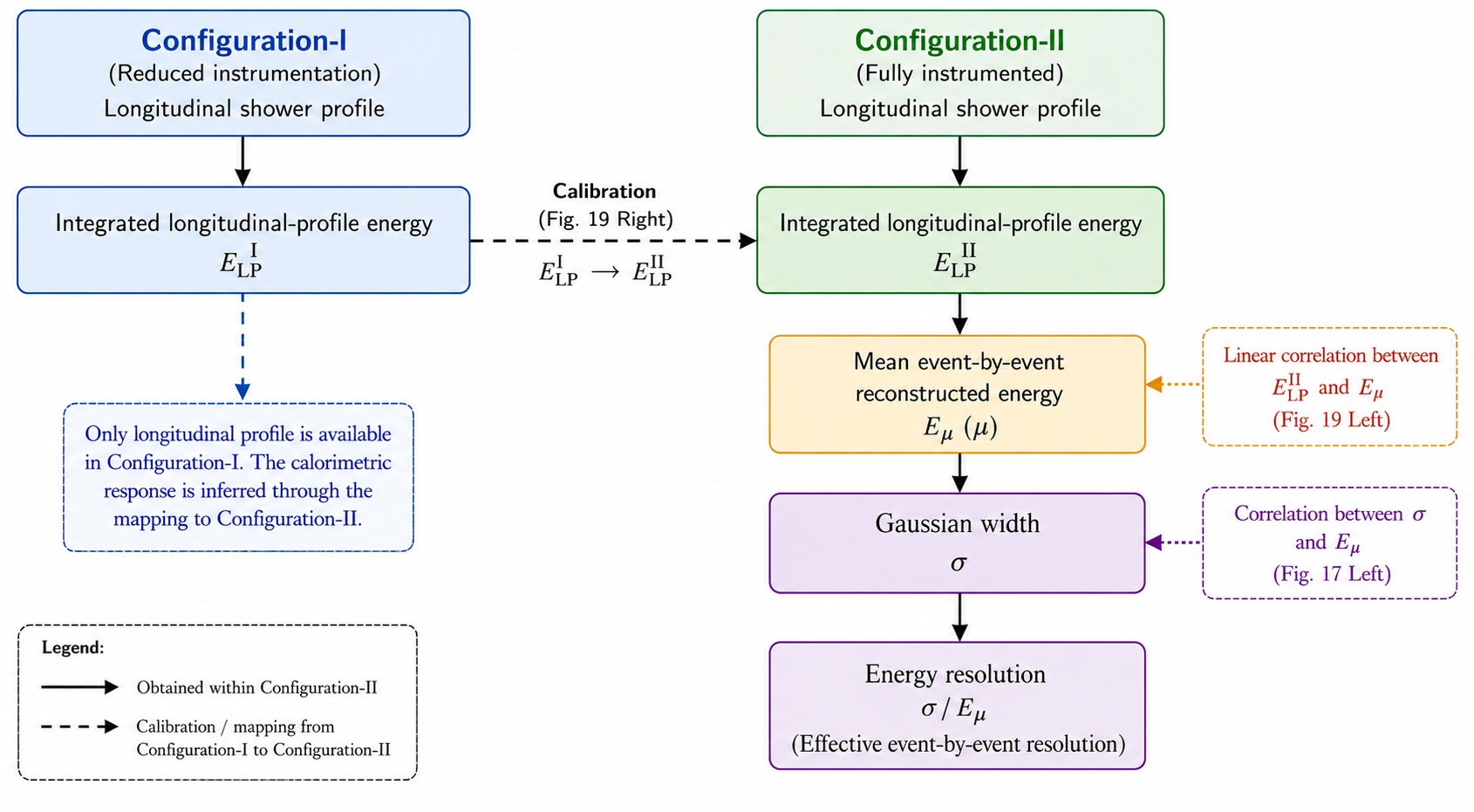}
    \caption{Schematic representation of the procedure adopted to estimate the effective event-by-event energy resolution of Configuration-I.}
    \label{fig:FlowChart}
\end{figure}

The agreement between the two reconstruction approaches is particularly significant for the present study. Since Configuration-I provides access only to the longitudinal shower profile, its calorimetric response must be inferred through the profile-integration method. To establish a correspondence between the two detector configurations, the reconstructed energies obtained from the integrated longitudinal shower profiles of Configurations I and II are compared in the right panel of Fig.~\ref{fig:13}. The observed linear relationship provides the calibration required to map the longitudinal-profile response of Configuration-I onto the equivalent response of Configuration-II. The calibrated longitudinal-profile energy is then converted to the corresponding mean event-by-event reconstructed energy using the linear correlation between the integrated longitudinal-profile energy and the mean reconstructed energy established for Configuration-II (left panel of Fig.~\ref{fig:13}). Finally, the relationship between the Gaussian width ($\sigma$) and the mean reconstructed energy, E$_{\mu}$, shown in Fig.~\ref{fig:11}, is used to determine the corresponding energy fluctuation. This procedure enables the effective energy resolution of Configuration-I to be estimated and directly compared with that of the fully instrumented calorimeter. A schematic illustration of the complete calibration and mapping procedure is presented in Fig.~\ref{fig:FlowChart}

\FloatBarrier

\begin{figure}[htbp]
    \centering
    \includegraphics[width=0.7\linewidth]{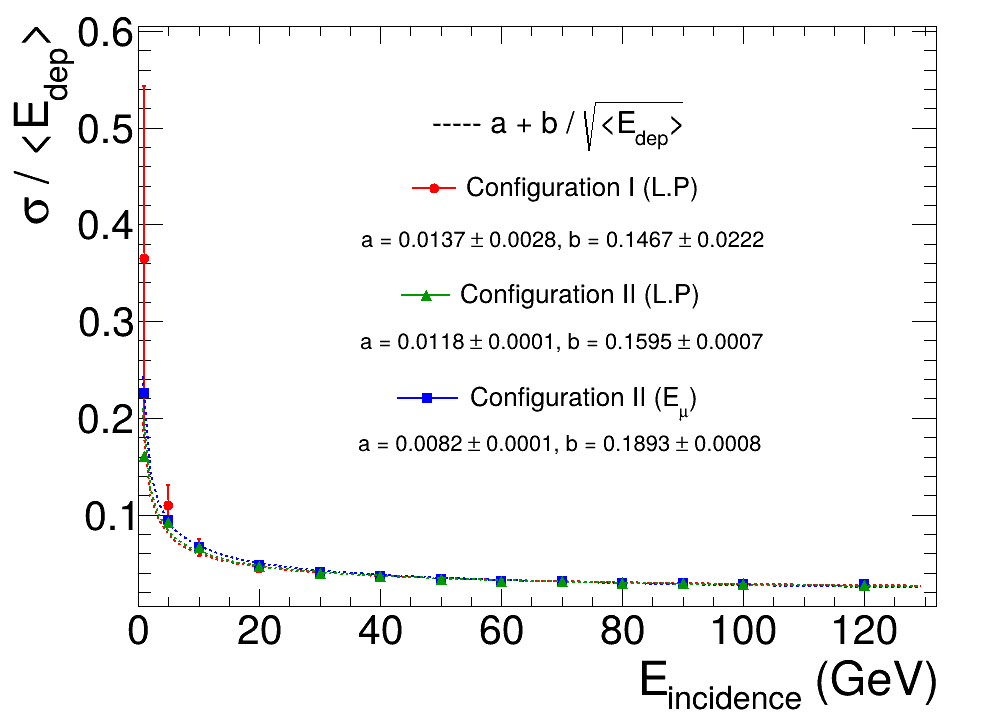}
    \caption{Comparison of the energy resolution ($\sigma/\langle E_{dep} \rangle$) for Configurations I and II as a function of the incident electron energy. For Configuration-II, $\langle E_{dep} \rangle$ is calculated in two approaches, i.e, by integrating longitudinal profile (L.P) and event-by-event Gaussian fits (E$_{\mu}$). The simulated data are fitted with the standard calorimetric resolution parameterization, Eq.~\ref{eq:resolution-II}. }

    \label{fig:17}
\end{figure}
\FloatBarrier

Figure~\ref{fig:17} compares the relative energy resolution, $\sigma/\langle E_{dep} \rangle$, obtained for Configurations I (red square) and II (blue square) as a function of the incident electron energy. The measured resolutions are parameterized using Eq.~(\ref{eq:resolution-II}). The fitted parameters for Configuration-I are a = 0.0137 and b = 0.1467, while the corresponding values for Configuration-II are a = 0.0082 and b = 0.1893. Although the individual fit parameters differ slightly, the resulting energy-resolution curves are found to be comparable throughout the investigated energy range, indicating that the reduced sequential-sampling prototype reproduces the principal calorimetric performance of the fully instrumented detector.

To further validate the longitudinal-profile reconstruction method, the energy resolution of Configuration-II was also determined by replacing the mean reconstructed energy obtained from the event-by-event Gaussian fits with the reconstructed energy obtained by integrating the longitudinal shower profile. These results are also superimposed in Fig.~\ref{fig:17} (green triangle). The two approaches yield nearly identical energy-resolution curves for Configuration-II, with fitted values of the parameters a and b in close agreement. This consistency demonstrates that the longitudinal-profile method provides a reliable estimate of the calorimetric response and justifies its application for estimating the effective event-by-event performance of Configuration-I.

For both configurations, the energy resolution improves with increasing incident energy, reflecting the reduction in relative shower fluctuations at higher energies. The measured resolutions are well described by Eq.~(\ref{eq:resolution-II}), yielding constant terms of approximately 1.4\% and 0.8\% for Configurations I and II, respectively. Although the constant term is conventionally associated with energy-independent contributions to the calorimeter resolution, the slightly larger value obtained for Configuration-I may arise from the sequential-sampling reconstruction procedure employed in the reduced prototype. Nevertheless, the close agreement between the measured resolution curves demonstrates that the reduced sequential-sampling prototype successfully reproduces the principal calorimetric performance of the fully instrumented detector. These results validate the use of Configuration-I as a practical and cost-effective approach for detector characterization and test-beam studies while substantially reducing the instrumentation requirements.

\section{Summary}

The performance of a silicon--tungsten electromagnetic calorimeter has been investigated using two detector Configurations: a reduced sequential-sampling prototype (Configuration-I) and a fully instrumented sampling calorimeter (Configuration-II). The response of the detector to minimum-ionizing particles and electromagnetic showers was first established through studies of energy deposition in individual silicon pads. A clear separation between pion-induced MIP signals and electron-induced shower signals was observed, demonstrating the particle-discrimination capability of the calorimeter.

For Configuration-I, the average longitudinal shower development was successfully reconstructed using measurements performed at different absorber depths. The resulting shower profiles were well described by a gamma-distribution parameterization, and the extracted shower maximum positions exhibited the expected logarithmic dependence on incident electron energy. The total calorimetric response obtained from the integrated longitudinal profiles showed excellent linearity with incident energy over the studied energy range. The effects of longitudinal leakage were found to be small at low energies and increased gradually with increasing beam energy, reaching approximately (3\%) at the highest simulated energies.

The fully instrumented geometry of Configuration-II enabled direct event-by-event reconstruction of the deposited energy. The reconstructed energy distributions were found to be well described by Gaussian functions, allowing the determination of both the mean calorimetric response and the corresponding energy resolution. In addition to reproducing the longitudinal shower characteristics observed in Configuration-I, the fully instrumented detector provided access to observables that cannot be measured directly in the reduced prototype, most notably event-by-event energy fluctuations.

A key outcome of the present study is the demonstration that the calorimetric response obtained from the integrated longitudinal shower profile exhibits a linear relationship with the response obtained from direct event-by-event energy reconstruction. This correspondence provides a calibration framework through which the response of Configuration-I can be mapped onto that of the fully instrumented detector. Consequently, an effective event-by-event calorimetric performance can be inferred for the reduced prototype, enabling a direct comparison of energy resolutions between the two configurations.

The comparison reveals that both detector configurations exhibit similar calorimetric behavior over the investigated energy range. The overall agreement in shower development, shower maximum position, linearity, and energy resolution indicates that the reduced sequential-sampling approach preserves the dominant characteristics of the fully instrumented calorimeter. These results demonstrate that a simplified detector configuration can provide a realistic representation of the calorimetric performance of a full silicon--tungsten calorimeter while significantly reducing instrumentation requirements.

\acknowledgments

Authors, S. Choudhury and M. Mondal gratefully acknowledge financial support from the DST-GOI under the scheme “Mega facilities in basic science research” [Sanction Order No. SR/MF/PS-02 2021-Jadavpur (E-37128) dated December 31, 2021]. One of the authors, A. Laha, acknowledges AICTE for financial support through AICTE Doctoral Fellowship (ADF), Government of India.

\bibliographystyle{unsrt}
\bibliography{biblio}

\end{document}